\documentstyle[aps,preprint]{revtex}
\begin{document}
\draft


\title{Faddeev-Jackiw Analysis of Topological Mass Generating Action}
\author{Chang-Yeong Lee$^{(a)}$\footnote{E-mail: leecy@phy.sejong.ac.kr}
 and Dong Won Lee$^{(b)}$\footnote{E-mail: theory@kkucc.konkuk.ac.kr}}
\address{ $^{(a)}$ Department of Physics,
                  Sejong University\\
                  Seoul 143-747, Korea \\
          $^{(b)}$ Department of Physics,
          Kon-kuk University\\
              Seoul 143-701, Korea
}
\maketitle

\begin{abstract}
 We analyze the gauge symmetry
of a topological mass generating action in four dimensions
which contains both
a vector and a second rank antisymmetric tensor
 fields. In the Abelian case, this system induces an effective mass
for the vector gauge field via a topological coupling $B \wedge F$ in
the presence of a kinetic term for the antisymmetric tensor field $B$,
while maintaining a gauge symmetry.
On the other hand, for the non-Abelian case the $B$ field does not
have a gauge symmetry unless an auxiliary vector field is introduced
to the system.
We analyze this change of symmetry in the Faddeev-Jackiw formalism,
and show how the auxiliary vector field enhances the symmetry.
At the same time this enhanced gauge symmetry becomes reducible.
We also show this phenomenon in this analysis.
\end{abstract}


\section{Introduction}
In 1991, there appeared a proposal that a vector field with Abelian
gauge symmetry in four dimensions can
develop an effective mass
 via a topological coupling with an antisymmetric tensor field,
while maintaining the symmetry \cite{abl}.
For the non-Abelian case, it was then shown that an auxiliary vector field
should be introduced to the system in order to have the same symmetry
property as in the Abelian case, that is both the vector and antisymmetric
tensor fields behave as gauge fields \cite{hl,al}.
Straightforward extension of the Abelian case to the non-Abelian one
does not work; no gauge symmetry for the antisymmetric tensor field.
In Ref. \cite{hl}, this was shown in the geometric BRST formalism.
There a clue for the understanding of this property came from the analysis
of the constraints among the equations of motion in both cases.

However, from the symmetry viewpoint this understanding is not quite enough.

In this paper, we analyze the symmetry property of this topological mass
 generating action in the Faddeev-Jackiw formalism.
Faddeev-Jackiw formalism \cite{fj,rj} is good for analyzing the symmetry
structure of a constrained system in the Hamiltonian formalism when
the Lagrangian is first order in time-derivatives.

To understand the Faddeev-Jackiw method, we now
consider a system of $N$ bosonic degrees of freedom, described by
the Lagrangian
\begin{equation}
L = a_k(q) \dot{q}_k -V(q), ~~ k=1, \dots, N.
\label{c1}
\end{equation}
Then, the equations of motion are given by
\begin{equation}
f_{ij}\dot{q}_j - \frac{\partial V}{\partial q_i} = 0
\label{c2}
\end{equation}
where the components of the symplectic two form
$f(q)= da(q)$ are given by
\begin{equation}
f_{ij}= \frac{\partial a_j}{\partial q_i} - \frac{\partial a_i}
    {\partial q_j} .
\label{c3}
\end{equation}
Here, $a=a_i dq_i$ is a canonical one form
whose components are given by the coefficients of $\dot{q}_k$ in
the Lagrangian (\ref{c1}).
If the symplectic matrix given by $f_{ij}$ is non-singular, then
its inverse matrix provides the values for the Dirac brackets of the
 theory \cite{cg}.
However, if the matrix $f_{ij}$ is singular, then
there will be constraints from the self consistency condition
 of the equations of
motion \cite{bnw}, which one can obtain by multiplying the
zero modes of the singular
matrix to the equations of motion Eq.(\ref{c2}):
\begin{equation}
\Omega^J \equiv (v_i^J)^T \frac{\partial V(q)}{\partial q_i} =0,
\label{c4}
\end{equation}
where the zero modes satisfy
\begin{equation}
(v_i^J)^T f_{ij} =0, ~~ J= 1, \dots, M,
\label{c5}
\end{equation}
and $M$ is the number of independent zero modes of $f_{ij}$.
There are two cases for consistency equations, Eq.(\ref{c4})
\cite{mw,hm,bb}.
The first case is when all the equations vanish identically.
This case corresponds to a theory with
 gauge symmetry.
In this  case one can simply choose a gauge and resolve the singularity.
The second case is when all or some of the equations give relations
between $q$'s. These relations among $q$'s are constraints, and one
needs to change
the Lagrangian into the following form to incorporate these constraints.
\begin{equation}
L = a_k(q) \dot{q}_k - \eta_J \Omega^J - V(q), ~~ k=1, \dots, N,
~~ J=1, \dots m, ~~ 0 < m \leq N .
\label{c6}
\end{equation}
Here $\eta_J$ are Lagrange multipliers.
The constraints should hold under time evolution and this can
be incorporated by putting the following constraints \cite{bnw,hm}
\[ \dot{\Omega}_J = 0, ~~ J = 1, \dots m, ~~ 0 < m \leq N, \]
which we implement by writing the Lagragian as
\begin{equation}
L = a_k(q) \dot{q}_k + \Omega^J \dot{\lambda}_J - V(q), ~~ k=1, \dots, N,
~~ J=1, \dots m, ~~ 0 < m \leq N .
\label{c7}
\end{equation}
Here we have
 changed the Lagrange multiplier field from $\eta_J$ to $\lambda_J$.
 Now, we have to check whether new constraints
arise or not from this new Lagrangian by repeating the above procedure,
regarding $q_k, \lambda_J$ as fields this time.
If the new symplectic matrix
is singular we repeat the whole procedure once again:
If all the consistency conditions for the equations of motion identically
vanish, thus having only the gauge symmetry, then
what we only have to do is a gauge fixing. The gauge fixing now makes
the symplectic matrix be nonsingular.
On the other hand, if new constraints for the fields $q_k, \lambda_J$ arise, 
then we have to repeat
the whole procedure once again. We have to repeat
this process until the symplectic matrix becomes nonsingular.
The first case happens when
the theory has only first class constraints in the Dirac formalism,
and the second case happens when the theory possesses
both first class (gauge symmetry) and second class constraints
in the Dirac formalism.
In this paper, we apply this method to analyze
 the symmetry of the topological mass generating action which contains
both a vector and an antisymmetric tensor fields.

So far, the antisymmetric tensor gauge theory
was analyzed by many in the Abelian case \cite{th,lhr,bs}.
In the non-Abelian case, however, the analysis of
the symmetry structure has not been done in the Hamiltonian formalism,
probably due to its complicated constraint structure.
The non-Abelian case was studied only in the geometric BRST
formalism \cite{tb,tn,ln,hl}, and we would like to analyze
the symmetry structure of the invariant action used in these works.

In Section II, we analyze the symmetry of the action with no auxiliary
vector field, and show that only the vector gauge field has non-Abelian
symmetry.
In Section III, we analyze the symmetry
after incorporating a vector auxiliary
field into the action, and show that both the vector and antisymmetric
 tensor fields have non-Abelian
gauge symmetry.
And in this case, the symmetry becomes reducible.
In Section IV, we conclude with discussions.

\section{Faddeev-Jackiw analysis of the action without a vector auxiliary
field}

We first start with the action extended
from the abelian case straightforwardly.
\begin{equation}
\int d^4 x {\cal L} = \int d^4 x {\rm Tr} \{ - \frac{1}{12}
        H_{\mu\nu\rho} H^{\mu\nu\rho} - \frac{1}{4} F_{\mu\nu} F^{\mu\nu} \}
\label{e1}
\end{equation}
where
\begin{eqnarray}
H_{\mu\nu\rho} = D_{[\mu} B_{\nu\rho]}
               = D_{\mu}B_{\nu\rho} +D_{\nu}B_{\rho\mu} +D_{\rho}B_{\mu \nu}
\label{e2}
\end{eqnarray}
and \( D_{\mu}B_{\nu\rho} = \partial_{\mu}B_{\nu\rho} +
[A_{\mu}, B_{\nu\rho}]. \)
Here, we dropped the $B \wedge F$ term from the action, since it does
not affect the result of the
analysis. The addition of $B \wedge F$ term only adds a few terms to
the constraints, but does not change the relations among constraints.
So, we drop it for convenience and briefness.
For the metric,
we use $g_{\mu \nu} = ( -, +, +, + )$ throughout the paper.

Introducing the conjugate momenta
\begin{eqnarray}
\Pi_{ij} & = & \dot{B}_{ij} + D_i B_{j0} - D_j B_{i0} + [A_0 ,B_{ij}],
\nonumber \\
\Pi_i & =  & 2( \dot{A}_i -D_i A_0 ),
\label{e3}
\end{eqnarray}
we can write the above Lagrangian in terms of conjugate momenta
\begin{equation}
{\cal L} = \frac{1}{4} \Pi_{ij}^a \dot{B}_{ij}^a
             + \frac{1}{4} \Pi_i^a \dot{A}_i^a - V_{(0)}
\label{e4}
\end{equation}
where
\begin{eqnarray}
V_{(0)}  &=& \frac{1}{2}\Pi_{ij}^a D_j B_{i0}^a - \frac{1}{4} \Pi_{ij}^a
             [A_0 , B_{ij}]^a + \frac{1}{8} \Pi_{ij}^2
            \nonumber \\
         & & \mbox{} + \frac{1}{4} \Pi_i^a D_i A_0^a + \frac{1}{16} \Pi_i^2
             + \frac{1}{8} F_{ij}^2 + \frac{1}{24} H_{ijk}^2.
\label{e5}
\end{eqnarray}
 From this Lagrangian we first get the components of the
canonical one form,
then we calculate a symplectic matrix with
 symplectic variables  $ B_{0i}^a  ,  B_{ij}^a  ,
\Pi_{ij}^a  ,   A_0^a  ,  A_i^a $ and $ \Pi_i^a $ 
(in order of appearance in the matrix).
With this symplectic matrix, we write a matrix equation for zero modes:
\begin{eqnarray}
\left [ \begin{array}{cccccc}
  0  & 0     & 0  &  0 & 0    & 0     \cr
  0 & 0          & P &         0 & 0          & 0     \cr
  0 & P^{\prime} & 0 &         0 & 0          & 0     \cr
  0 & 0          & 0 &         0 & 0          & 0     \cr
  0 & 0          & 0 &         0 & 0          & T     \cr
  0 & 0          & 0 &         0 & T^{\prime} & 0
   \end{array} \right ]
 \left[ \begin{array}{c}
    \alpha_l^g     \cr
    \beta_{lm}^g   \cr
    \gamma_{lm}^g  \cr
    \rho_0^g       \cr
    \sigma_l^g     \cr
    \kappa_l^g
    \end{array} \right]
= 0
\end{eqnarray}
where
\begin{eqnarray*}
P & \equiv & f^{(0) B \Pi}_{~~ija~lmg} = -\frac{1}{4} \delta^{ag}
      \delta_{ij}^{lm} \delta ({\bf x} - {\bf y}) \\
P' & \equiv & f^{(0) \Pi B}_{~~ija~lmg} = \frac{1}{4} \delta^{ag}
    \delta_{ij}^{lm} \delta ({\bf x} - {\bf y}) \\
T & \equiv & f^{(0) A \Pi}_{~~ia~lg}   = \mbox{} -\frac{1}{4}
    \delta_{il} \delta^{ag} \delta ({\bf x}-{\bf y}) \\
T^{\prime} & \equiv & f^{(0) \Pi A}_{~~ia~lg}  = \frac{1}{4} \delta_{il}
    \delta^{ag} \delta ({\bf x}-{\bf y}) .
\end{eqnarray*}
Throughout the paper it will be understood that all quantities are taken
at equal time. The above symplectic matrix is singular because
there exist nontrivial eigenvectors with zero eigenvalue.
 Now we can write new constraints from the zero modes
as
\begin{eqnarray}
\Omega^{(0)} &=& \int d^3 {\bf x} \{~\alpha_l^g ({\bf x}) \frac{\delta}
                {\delta B_{0l}^g ({\bf x})} +\rho_0^g ({\bf x}) \frac{\delta}
                {\delta A_0^g ({\bf x})} \} \int d^3 {\bf y}
                V_{(0)} .
                \label{e7}
\end{eqnarray}
 Since $\alpha_l^g ({\bf x})$ and $\rho_0^g ({\bf x})$ are arbitrary
parameters, we write the constraints and their Lagrange multipliers 
as follows.
\begin{eqnarray}
\Omega^{(0)}_1 &=& D_j \Pi_{ji}^a ~~~~,~~~\eta_i^a  \nonumber \\
\Omega^{(0)}_2 &=& D_i \Pi_i^a + [B_{ij} ,\Pi_{ij}]^a ~,~\omega^a  .
\label{e8}
\end{eqnarray}
 Incorporating these new constraints, Lagrangian now becomes
\begin{eqnarray}
{\cal L} &=& \frac{1}{4} \Pi_{ij}^a \dot{B}_{ij}^a + \frac{1}{4} \Pi_i^a
             \dot{A}_i^a +(D_j \Pi_{ji})^a \dot{\eta}_i^a
             \label{e9} \\
         & & \mbox{} + (D_i \Pi_i + [B_{ij} ,\Pi_{ij}])^a
       \dot{\omega}^a - V_{(1 )} \nonumber
\end{eqnarray}
where
\begin{eqnarray}
V_{(1)}  &=& \frac{1}{8} \Pi_{ij}^2 +\frac{1}{16} \Pi_i^2 +\frac{1}{8} F_{ij}^2
             +\frac{1}{24} H_{ijk}^2 .
\label{e10}
\end{eqnarray}
Repeating the same procedure, we obtain a new symplectic matrix,
and write a matrix equation for zero modes as follows.
\begin{eqnarray}
\left [ \begin{array}{cccccc}
  0 & P & 0 & 0 & 0 & R                 \cr
  P^{\prime} & 0 & 0 & 0 & S & U        \cr
  0 & 0 & 0 & T & V & W                 \cr
  0 & 0 & T^{\prime} & 0 & 0 & X                       \cr
  0 & S^{\prime} & V^{\prime} & 0 & 0 & 0              \cr
  R^{\prime} & U^{\prime} & W^{\prime} & X^{\prime} & 0 & 0  \cr
    \end{array} \right ]
 \left [ \begin{array}{c}
    \alpha_{lm}^g  \cr
    \beta_{lm}^g   \cr
    \gamma_l^g     \cr
    \rho_l^g       \cr
  \sigma_l^g     \cr
    \nu^g          \cr
    \end{array} \right ]
= 0
\end{eqnarray}
where
\begin{eqnarray}
 R^{\prime} &\equiv& f^{gac}\Pi_{ij}^c \delta({\bf x} - {\bf y})
                               ~,~~
 R^{\prime} \equiv \mbox{} -f^{agc}\Pi_{lm}^c \delta ({\bf x}
 - {\bf y})            \cr
 S &\equiv& D_m^y \delta^{ag}\delta_{ij}^{ml}
                      \delta ({\bf x} - {\bf y})
                               ~,~~
 S^{\prime} \equiv \mbox{} -D_j^x \delta^{ag}\delta_{ji}^{lm}
                                \delta ({\bf x} - {\bf y})
                           \cr
 U &\equiv& f^{gba}B_{ij}^b \delta ({\bf x} - {\bf y})
                ~,~~
 U^{\prime} \equiv \mbox{} -f^{abg}B_{lm}^b \delta ({\bf x}
    - {\bf y})              \cr
 V &\equiv& f^{gac}\Pi_{il}^c \delta ({\bf x} - {\bf y})
                        ~,~~
 V^{\prime} \equiv \mbox{} -f^{agc}\Pi_{li}^c \delta ({\bf x}
    - {\bf y })                \cr
 W &\equiv& f^{gac}\Pi_i^c \delta ({\bf x} - {\bf y})
                    ~,~~
 W^{\prime} \equiv \mbox{} -f^{agc}\Pi_l^c \delta ({\bf x}
   - {\bf y})                \cr
 X &\equiv& \mbox{} -D_i \delta^{ag}  \delta ({\bf x}
   - {\bf y})           ~,~~
 X^{\prime} \equiv \mbox{} -D_l \delta^{ag}  \delta ({\bf x}
    - {\bf y})    \nonumber
\end{eqnarray}
Here, symplectic variables are $ B_{ij}^a , \Pi_{ij}^a , A_i^a ,
 \Pi_i^a , \eta_i^a $ and $ \omega^a $ in order of appearance in
the symplectic matrix.
 From the matrix equation above, 
we find two zero modes with
independent parameters $\sigma_l$ and $\nu$:
\begin{eqnarray}
(\alpha_{lm} = 4 [B_{lm}, \nu], ~\beta_{lm} = 4[\Pi_{lm}, \nu],
 ~ \gamma_l= 4 D_i \nu, ~\rho_l= 4[\Pi_{i}, \nu], ~\sigma_l ,
 ~\nu)
\label{e12} .
\end{eqnarray}
Among these two zero modes,
only the zero mode with $\sigma_l$ provides
a new constraint
\begin{eqnarray}
\Omega^{(1)}_1 = [F_{jk}, H_{ijk}] -[\Pi_j, \Pi_{ji}].
\label{e13}
\end{eqnarray}
The consistency condition from the zero mode related  to  $\nu$
 vanishes identically.
 Thus  new Lagrangian is given by 
\begin{eqnarray}
{\cal L} &=& \frac{1}{4} \Pi_{ij}^a \dot{B}_{ij}^a + \frac{1}{4} \Pi_i^a
             \dot{A}_i^a +(D_j \Pi_{ji})^a \dot{\eta}_i^a
          \mbox{} + (D_i \Pi_i + [B_{ij} ,\Pi_{ij}])^a \dot{\omega}^a \cr
         & & \mbox{} + ([F_{jk}, H_{ijk}] -[\Pi_j, \Pi_{ji}])^a
   \dot{\xi}_i^a - V_{(2)}
\label{e14}
\end{eqnarray}
where
\begin{eqnarray}
V_{(2)}  &=& \frac{1}{8} \Pi_{ij}^2 +\frac{1}{16} \Pi_i^2 +\frac{1}{8}
  F_{ij}^2 +\frac{1}{24} H_{ijk}^2 \mid_{\Omega_1^{(1)}}
\label{e15} .
\end{eqnarray}
Here the symplectic variable $ \xi_i^a $ is added.
Then, the symplectic matrix and its zero mode equation is 
\begin{eqnarray}
\left [ \begin{array}{ccccccc}
  0 & P & 0 & 0 & 0 & R & \Psi               \cr
  P^{\prime} & 0 & 0 & 0 & S & U & \Sigma       \cr
  0 & 0 & 0 & T & V & W & \phi                \cr
  0 & 0 & T^{\prime} & 0 & 0 & X & \chi                      \cr
  0 & S^{\prime} & V^{\prime} & 0 & 0 & 0 & 0             \cr
  R^{\prime} & U^{\prime} & W^{\prime} & X^{\prime} & 0 & 0 & 0 \cr
  \Psi^{\prime} & \Sigma^{\prime} & \phi^{\prime} & \chi^{\prime} & 0 & 0
  & 0 \cr
    \end{array} \right ]
 \left [ \begin{array}{c}
    \alpha_{lm}^g  \cr
    \beta_{lm}^g   \cr
    \gamma_l^g     \cr
    \rho_l^g       \cr
    \sigma_l^g     \cr
    \nu^g          \cr
    \mu_l^g        \cr
    \end{array} \right ]
=0
\end{eqnarray}
where
\begin{eqnarray*}
 \Psi & = & f^{gbc}F_{mn}^b \{(\partial_l^y\delta^{ca}
             + f^{cda}A_l^d ) \delta_{ij}^{mn} \\
         & & \mbox{} + (\partial_m^y\delta^{ca} + f^{cda}A_m^d )
  \delta_{ij}^{nl}
             + (\partial_n^y\delta^{ca} + f^{cda}A_n^d ) \delta_{ij}^{lm}
             \}\delta ({\bf x} - {\bf y}), \\
 \Psi' & = & \mbox{} -f^{abc}F_{jk}^b \{(\partial_i^x\delta^{cg}
             + f^{cdg}A_i^d ) \delta_{lm}^{jk} \\
         & & \mbox{} + (\partial_j^x\delta^{cg} + f^{cdg}A_j^d )
    \delta_{lm}^{ki}
             + (\partial_k^x\delta^{cg} + f^{cdg}A_k^d ) \delta_{lm}^{ij}
             \}\delta ({\bf x} - {\bf y}),  \\
 \phi & = & f^{gbc} \{ (2\partial_j^y \delta^{ba}
             \delta_{ik} +f^{bae}\delta_{ij}A_k^e + f^{bda}\delta_{ki}
             A_j^d )\delta ({\bf x} - {\bf y}) \} H_{ljk}^c \\
         & & \mbox{} +f^{gbc}f^{cae}F_{jk}^b (\delta_{li}B_{jk}^e +\delta_{ji}
             B_{kl}^e +\delta_{ki}B_{lj}^e)\delta ({\bf x} - {\bf y}), \\
 \phi' & = & \mbox{} -f^{abc} \{ (2\partial_j^x \delta^{bg}
           \delta_{kl} +f^{bge}\delta_{jl}A_k^e + f^{bdg}\delta_{kl}
             A_j^d )\delta ({\bf x} - {\bf y}) \} H_{ijk}^c \\
         & & \mbox{} -f^{abc}f^{cge}F_{jk}^b (\delta_{li}B_{jk}^e +\delta_{jl}
             B_{ki}^e +\delta_{kl}B_{ij}^e)\delta ({\bf x} - {\bf y}), \\
 \Sigma & = & \mbox{} -f^{gba}\Pi_m^b \delta_{ij}^{ml}
             \delta ({\bf x} - {\bf y}), \\
  \Sigma' & = & f^{abg}\Pi_j^b \delta_{ji}^{lm}
             \delta ({\bf x} - {\bf y}), \\
  \chi & = & \mbox{} -f^{gac}\Pi_{il}^c \delta ({\bf x} i
     - {\bf y}), \\
 \chi' & = & f^{agc}\Pi_{li}^c \delta ({\bf x} - {\bf y}),
\end{eqnarray*}
and $ P, P', R, $ etc. are the same as before.
Again this symplectic matrix is singular,
and after solving the zero mode equation
 we find a zero mode: 
\begin{equation}
(~\alpha_{ij}=4[B_{ij}, \nu] , ~\beta_{ij}=4[\Pi_{ij}, \nu],
 ~ \gamma_i = 4 D_i \nu, ~ \rho_i = 4[\Pi_{i}, \nu],
   ~  \sigma_i=0, ~ \nu, ~\mu_i =0).
\label{e17}
\end{equation}
With this zero mode, we see that the constraint equation vanishes
identically:
\begin{eqnarray}
\Omega^{(2)} &=&  \int d^3 {\bf x} \{4[B_{lm},\nu]^g \frac{\delta}
                 {\delta B_{lm}^g} +4[\Pi_{lm}, \nu]^g \frac{\delta}
                 {\delta \Pi_{lm}^g} \cr
             & & \mbox{} +4[\Pi_l, \nu]^g \frac{\delta}{\delta \Pi_l^g }
                 +4D_l\nu^g \frac{\delta}{\delta A_l^g} \}
                 \int V_{(1)} d^3 {\bf y} \cr
  & \equiv & 0 . \nonumber
\end{eqnarray}
This shows that the theory we are considering has gauge symmetry and
the gauge transformation is given by the above zero mode.
Namely, the gauge transformations of the fields are given by
 $\delta B_{ij} =\alpha_{ij} = 4[B_{ij}, \nu],
 ~ \delta A_i =  \gamma_i = 4 D_i \nu . $
This clearly shows that only the vector field has non-Abelian gauge
symmetry, unlike the Abelian case \cite{abl}
 where both the vector and antisymmetric
tensor fields behave as gauge fields. 

 Finally, to remove the singularity due to the above gauge symmetry,
 we choose a gauge as
\begin{eqnarray}
\partial_i A_i = 0.
\end{eqnarray}
Then the Lagrangian becomes
\begin{eqnarray}
{\cal L} &=& \frac{1}{4} \Pi_{ij}^a \dot{B}_{ij}^a + \frac{1}{4} \Pi_i^a
             \dot{A}_i^a +(D_j \Pi_{ji})^a \dot{\eta}_i^a
          + (D_i \Pi_i + [B_{ij} ,\Pi_{ij}])^a \dot{\omega}^a \cr
         & & \mbox{} + ([F_{jk}, H_{ijk}] -[\Pi_j, \Pi_{ji}])^a \dot{\xi}_i^a
             + (\partial_i A_i^a) \dot{\lambda}^a - V_{(3)}
\label{e18}
\end{eqnarray}
where
\begin{eqnarray*}
V_{(3)}  =  V_{(2)} \mid_{\partial_i A_i = 0}.
\end{eqnarray*}
Now, the symplectic matrix with an added symplectic variable $ \lambda^a $
is given by
\begin{eqnarray}
\left [ \begin{array}{cccccccc}
  0 & P & 0 & 0 & 0 & R & \Psi & 0              \cr
  P^{\prime} & 0 & 0 & 0 & S & U & \Sigma & 0      \cr
  0 & 0 & 0 & T & V & W & \phi & Y              \cr
  0 & 0 & T^{\prime} & 0 & 0 & X & \chi & 0                    \cr
  0 & S^{\prime} & V^{\prime} & 0 & 0 & 0 & 0 & 0           \cr
  R^{\prime} & U^{\prime} & W^{\prime} & X^{\prime} & 0 & 0 & 0 & 0 \cr
  \Psi^{\prime} & \Sigma^{\prime} & \phi^{\prime} & \chi^{\prime} & 0 & 0
 & 0 & 0 \cr
  0 & 0 & Y^{\prime} & 0 & 0 & 0 & 0 & 0      \cr
    \end{array} \right ]
\end{eqnarray}
where
\begin{eqnarray}
 Y  =   -\partial_i \delta ({\bf x} - {\bf y})
            \delta^{ag}, ~~~~~
 Y'  =   -\partial_l \delta ({\bf x} - {\bf y})
            \delta^{ag} \nonumber ,
\end{eqnarray}
and $ P, P', R, $ etc. are the same as before. One can check
that this symplectic matrix is nonsingular, as it should be.


\section{Faddeev-Jackiw analysis of the action with a vector auxiliary
field}

In the previous section, we have seen that the straightforward extension of
the Abelian action to the non-Abelian one does not work.
Thus, following Refs. \cite{tb,tn,hl}, we introduce an auxiliary vector field
to the theory by replacing
$ B_{\mu \nu} \longrightarrow  B_{\mu \nu} - D_{ [\mu} K_{\nu ]} $,
where $K_{\nu}$ is an auxiliary vector field.
This replacement also changes the field strength of the antisymmetric
tensor field into 
\begin{equation}
 H_{\mu\nu\rho} = D_{[\mu} B_{\nu\rho]} \longrightarrow
  {H'}_{\mu\nu\rho} =  D_{[\mu} B_{\nu\rho]} -
                   [F_{[\mu\nu},K_{\rho]}] .
\label{f2}
\end{equation}
Now we write the Lagrangian with this new field strength $H'$
\begin{eqnarray}
{\cal L} &=&  {\rm Tr} \{ - \frac{1}{12} {H'}_{\mu\nu\rho} {H'}^{\mu\nu\rho}
                          - \frac{1}{4} F_{\mu\nu} F^{\mu\nu} \} 
                          \label{f3}.
\end{eqnarray}
Introducing the canonical momenta
\begin{eqnarray}
\Pi_{ij} &=& \frac{1}{2} (\dot{B}_{ij} + D_i B_{j0} - D_j B_{i0}
               + [A_0 ,B_{ij}] -[\dot{A}_i , K_j] \nonumber \\
           & & \mbox{} +[\dot{A}_j ,K_i] + [D_i A_0 , K_j] -[D_j A_0 , K_i]
               -[F_{ij} , K_0] ),
           \label{f4} \\
\Pi_i & = &  \dot{A}_i -D_i A_0 ,
\nonumber
\end{eqnarray}
we rewrite the Lagrangian in its first order form
\begin{eqnarray}
{\cal L} &=& \frac{1}{2} \Pi_{ij}^a \dot{B}_{ij}^a + \frac{1}{2} ([A_i ,K_j]
             -[A_j ,K_i ])^a \dot{\Pi}_{ij}^a \nonumber \\
         & & \mbox{} +[A_j , \Pi_{ij}]^a \dot{K}_i^a + \frac{1}{2} \Pi_i^a
      \dot{ A}_i^a - V_{(0)} \label{f6}
\end{eqnarray}
where
\begin{eqnarray}
V_{(0)}  &=& \Pi_{ij}^a D_j B_{i0}^a - \frac{1}{2} \Pi_{ij}^a [A_0 , B_{ij}]^a
             +\Pi_{ij}^a [D_j A_0 , K_i] \nonumber \\
         & & \mbox{} + \frac{1}{2} \Pi_{ij}^a [F_{ij} , K_0]^a +
   \frac{1}{2} \Pi_{ij}^2
             + \frac{1}{2} \Pi_i^a D_i A_0^a \nonumber \\
         & & \mbox{} + \frac{1}{4} \Pi_i^2 + \frac{1}{8} F_{ij}^2
             + \frac{1}{24} H_{ijk}^2 \nonumber .
\end{eqnarray}
Repeat the procedure in the previous section, we first obtain
a zero mode equation for
the symplectic matrix
\begin{eqnarray}
\left [ \begin{array}{cccccccc}
  0 & 0          & 0 &         0 & 0          & 0& 0 & 0    \cr
  0 & 0          & P &         0 & 0          & 0& 0 & 0    \cr
  0 & P^{\prime} & 0 &         0 & Q          & 0& 0 & 0    \cr
  0 & 0          & 0 &         0 & 0          & 0& 0 & 0    \cr
  0 & 0          & Q^{\prime} &0 & 0          & S& 0 & T    \cr
 0 & 0          & 0 &         0 & S^{\prime} & 0& 0 & 0    \cr
  0 & 0          & 0 &         0 & 0          & 0& 0 & 0    \cr
  0 & 0          & 0 &         0 & T^{\prime} & 0& 0 & 0
    \end{array} \right ]
 \left [ \begin{array}{c}
    \alpha_l^g     \cr
    \beta_{lm}^g   \cr
    \gamma_{lm}^g  \cr
    \rho_0^g       \cr
    \sigma_l^g     \cr
    \phi_l^g       \cr
    \mu_0^g        \cr
    \nu_l^g
    \end{array} \right ]
= 0
\end{eqnarray}
where
\begin{eqnarray*}
  P & = &  -\frac{1}{2} \delta^{ag} \delta_{ij}^{lm}
                             \delta ({\bf x} - {\bf y}) \\
 P' & = & \frac{1}{2}\delta^{ag} \delta_{lm}^{ij}
                              \delta ({\bf x} - {\bf y}) \\
 Q & = & \mbox{} -\frac{1}{2}f^{abc} \delta^{bg} (\delta_{il}K_j^c
                  - \delta_{jl} K_i^c ) \delta({\bf x}-{\bf y}) \\
 Q'  & = & \frac{1}{2}f^{gbc} \delta^{ba} (\delta_{il}K_m^c
                           -\delta_{mi} K_l^c) \delta({\bf x}-{\bf y})\\
 S   & = & \mbox{} -\frac{1}{2} \delta_{il} \delta^{ag}
                              \delta ({\bf x}-{\bf y}) \\
 S'   &  = & \frac{1}{2} \delta_{il} \delta^{ag}
                              \delta ({\bf x}-{\bf y}) \\
 T   & = &  f^{gbc} \delta^{ba} \Pi_{li}^c
                              \delta ({\bf x}-{\bf y}) \\
 T'   & = & \mbox{} - f^{abc} \delta^{bg} \Pi_{il}^c
                              \delta ({\bf x}-{\bf y}) .
\end{eqnarray*}
Here the symplectic variables are $ B_{0i}^a , B_{ij}^a , \Pi_{ij}^a ,
 A_0^a , A_i^a , \Pi_i^a , K_0^a $ and $ K_i^a  $ in order of 
appearance in the symplectic matrix.
 The matrix equation has four zero modes with four independent variables
$\alpha_l, ~ \rho_0, ~ \mu_0, ~ \nu_l$:
\begin{eqnarray}
(~\alpha_l , ~0, ~0, ~\rho_0 , ~0, ~2[\Pi_{ml} , \nu_m] , ~\mu_0
 , ~\nu_l ),
\end{eqnarray}
These zero modes yield four constraints, and we write them
with their respective Lagrange multiplier below.
\begin{eqnarray}
\Omega^{(0)}_1 &=& D_j \Pi_{ji}^a ;~~~\eta_i^a  \cr
\Omega^{(0)}_2 &=& D_i \Pi_i^a + [B_{ij} ,\Pi_{ij}]^a -
  2[\Pi_{ij} ,D_j K_i]^a ;~~\omega^a  \cr
\Omega^{(0)}_3 &=& [F_{ij} ,\Pi_{ij}]^a ;~~~\theta^a \label{f10} \\
\Omega^{(0)}_4 &=& [\Pi_{ji} ,\Pi_j]^a -\frac{1}{4}[H_{ijk} ,F_{jk}]^a;
                  ~~\chi_i^a  \nonumber .
\end{eqnarray}
However, these four constraints are not all independent.
The first and third constraints are related by the following
equation.
\begin{eqnarray}
D_i \Omega_1^{(0)} + \frac{1}{2} \Omega_3^{(0)} = 0 .
\label{f7}
\end{eqnarray}
In order to incorporate this dependence between the two constraints, 
we further
introduce a new constraint and its Lagrange multiplier 
\begin{eqnarray}
\Omega^{(0)}_5 &=& D_i \eta_i^a + \frac{1}{2} \theta^a   ;~~\lambda^a
\label{f8} ,
\end{eqnarray}
and write the Lagrangian as
\begin{eqnarray}
{\cal L} &=& \frac{1}{2} \Pi_{ij}^a \dot{B}_{ij}^a + \frac{1}{2} ([A_i ,K_j]
             -[A_j ,K_i ])^a \dot{\Pi}_{ij}^a \cr
         & & \mbox{} +[A_j , \Pi_{ij}]^a \dot{K}_i^a + \frac{1}{2}
      \Pi_i^a \dot{A}_i^a
             +(D_j \Pi_{ji})^a \dot{\eta}_i^a \cr
   & & \mbox{} + (D_i \Pi_i + [B_{ij} ,\Pi_{ij}] - 2[\Pi_{ij} ,D_j K_i])^a
             \dot{\omega}^a  \label{f9} \\
     & & \mbox{} +[F_{ij} ,\Pi_{ij}]^a \dot{\theta}^a + ([\Pi_{ji} ,\Pi_j]
              -\frac{1}{4}[H_{ijk} ,F_{jk}])^a \dot{\chi}_i^a \cr
    & & \mbox{} + (D_i \eta_i^a + \frac{1}{2} \theta^a )\dot{\lambda}^a
      - V_{(1)} \nonumber
\end{eqnarray}
where
\begin{eqnarray}
V_{(1)}  &=& \frac{1}{2} \Pi_{ij}^2 +\frac{1}{4} \Pi_i^2 +\frac{1}{8} F_{ij}^2
             +\frac{1}{8} H_{ijk}^a D_i B_{jk}^a \cr
         & & + \frac{1}{2} K_i^a [\Pi_{ij} ,\Pi_j ]^a .
\end{eqnarray}
With the symplectic variables  $B_{ij}^a , \Pi_{ij}^a , A_i^a ,\Pi_i^a ,
K_i^a , \eta_i^a , \omega^a , \theta^a , \chi_i^a $ and $ \lambda^a $ 
(in order of appearance in the symplectic matrix), 
we obtain the following zero mode equation for the symplectic matrix.
\begin{eqnarray}
\left [ \begin{array}{cccccccccc}
  0 & P & 0 & 0 & 0 & 0 & C & 0 & D & 0                \cr
  P^{\prime} & 0 & Q & 0 & 0 & E & F & G & H & 0       \cr
  0 & Q^{\prime} & 0 & S & T & I & J & K & L & M       \cr
  0 & 0 & S^{\prime} & 0 & 0 & 0 & N & 0 & A & 0       \cr
  0 & 0 & T^{\prime} & 0 & 0 & 0 & B & 0 & U & 0       \cr
  0 & E^{\prime} & I^{\prime} & 0 & 0 & 0 & 0 & 0 & 0 & V    \cr
  C^{\prime} & F^{\prime} & J^{\prime} & N^{\prime} & B^{\prime} & 0 & 0
      & 0 & 0 & 0      \cr
  0 & G^{\prime} & K^{\prime} & 0 & 0 & 0 & 0 & 0& 0 & W     \cr
  D^{\prime} & H^{\prime} & L^{\prime} & A^{\prime} & U^{\prime} & 0
      & 0 & 0 & 0 & 0  \cr
  0 & 0 & M^{\prime} & 0 & 0 & V^{\prime} & 0 & W^{\prime} & 0 & 0
    \end{array} \right ]
 \left [ \begin{array}{c}
    \alpha_{lm}^g  \cr
    \beta_{lm}^g   \cr
    \gamma_l^g     \cr
    \rho_l^g       \cr
    \sigma_l^g     \cr
    \mu_l^g        \cr
    \nu^g          \cr
  \xi^g          \cr
    \psi_l^g       \cr
    \phi^g
    \end{array} \right ]
= 0 \label{m39}
\end{eqnarray}
where
\begin{eqnarray*}
 C &=& f^{gac}\Pi_{ij}^c \delta({\bf x} - {\bf y})
                               \\
C' &=& \mbox{} -f^{agc}\Pi_{lm}^c \delta ({\bf x} - {\bf y})
\\
 D &=& \frac{1}{8}f^{gbc}F_{mn}^b \{(\partial_l^y
                \delta^{ca} +f^{cde}A_l^d\delta^{ea})\delta_{ij}^{mn}
                +(\partial_m^y \delta^{ca} +f^{cde}A_m^d\delta^{ea}
             )\delta_{ij}^{nl} \\
           & &  \mbox{} + (\partial_n^y \delta^{ca} +f^{cde}A_n^d
                \delta^{ea})\delta_{ij}^{lm} \}\delta ({\bf x} - {\bf y})
                \\
 D' &=& \mbox{} -\frac{1}{8}f^{abc}F_{jk}^b \{(\partial_i^x
                \delta^{cg} +f^{cde}A_i^d\delta^{eg})\delta_{jk}^{lm}
                +(\partial_j^x \delta^{cg} +f^{cde}A_j^d\delta^{eg}
                )\delta_{ki}^{lm} \\
           & &  \mbox{} + (\partial_k^x \delta^{cg} +f^{cde}A_k^d
                \delta^{eg})\delta_{ij}^{lm} \}\delta ({\bf x} - {\bf y})
                \\
 E &=& \frac{1}{2}(\partial_m^y \delta^{ag}
                +f^{gbc}A_m^b \delta^{ca})\delta_{ij}^{ml}
                \delta ({\bf x} - {\bf y})  \\
 E' &=& \mbox{} -\frac{1}{2}(\partial_j^x \delta^{ag}
                +f^{abc}A_j^b \delta^{cg})\delta_{ij}^{ml}
                \delta ({\bf x} - {\bf y})  \\
 F &=& (f^{gba}B_{ij}^b +2f^{gac}(D_iK_j^c
              -D_jK_i^c)) \delta ({\bf x} - {\bf y})
                \\
 F' &=& \mbox{} -(f^{abg}B_{lm}^b +2f^{agc}(D_lK_m^c
                -D_mK_l^c)) \delta ({\bf x} - {\bf y})
                \\
 G &=& f^{gba}F_{ij}^b \delta ({\bf x} - {\bf y})
                \\
 G' &=& \mbox{} -f^{abg}F_{lm}^b \delta ({\bf x} - {\bf y})
                \\
 H &=& \frac{1}{2} f^{gac}\Pi_m^c \delta_{ij}^{ml}
                \delta ({\bf x} - {\bf y})   \\
 H' &=& \mbox{} -\frac{1}{2} f^{agc}\Pi_j^c \delta_{ji}^{lm}
                \delta ({\bf x} - {\bf y})    \\
 I &=& f^{gac}\Pi_{il}^c \delta ({\bf x} - {\bf y})
               \\
 I' &=& \mbox{} -f^{agc}\Pi_{li}^c \delta ({\bf x} - {\bf y})
                \\
 J &=& f^{gac}\Pi_i^c -2 f^{gbc}f^{cae}\Pi_{li}^bK_l^e
                \delta ({\bf x} - {\bf y}) \\
 J' &=& -f^{agc}\Pi_l^c +2 f^{abc}f^{cge}\Pi_{il}K_i^e
                \delta ({\bf x} - {\bf y})
                \\
 K &=& 2f^{gbc} \{(\partial_l^y \delta^{ba}
                +f^{bda}A_l^d) \delta ({\bf x} - {\bf y}) \}\Pi_{li}^c
                \\
 K' &=& \mbox{} -2f^{abc} \{(\partial_i^x \delta^{bg}
                +f^{bdg}A_i^d) \delta ({\bf x} - {\bf y}) \}\Pi_{il}^c
                \\
 L &=& \frac{1}{2}f^{gbc} \{(\partial_j^y \delta^{ba}
                +f^{bda}A_j^d) \delta ({\bf x} - {\bf y}) \} H_{lji}^c \\
    &  &  \mbox{} +\frac{1}{4}f^{gbc}F_{jk}^b \{f^{cae}(\delta_{il}B_{jk}^e
             +\delta_{ji}B_{kl}^e +\delta_{ki}B_{lj}^e)
                \delta ({\bf x} - {\bf y}) \\
            & & \mbox{} -f^{cde} ((\partial_l^y \delta_{ji}\delta^{da}
                -\partial_j^y \delta_{li} \delta^{da} +f^{dah}
                \delta_{li}A_j^h +f^{dfa}\delta_{ji}
                A_l^f)\delta ({\bf x} - {\bf y}))K_k^e \\
            & & \mbox{} -f^{cde}((\partial_j^y \delta_{ki}\delta^{da}
         -\partial_k^y
                \delta_{ji}\delta^{da} +f^{dah} \delta_{ji}A_k^h +f^{dfa}
                \delta_{ki}A_j^f)\delta ({\bf x} - {\bf y}))K_l^e \\
            & & \mbox{} -f^{cde}((\partial_k^y \delta_{li}\delta^{da}
        -\partial_l^y
                \delta_{ki} \delta^{da} +f^{dah} \delta_{ki}A_l^h
                +f^{dfa}\delta_{li} A_k^f)\delta ({\bf x} - {\bf y}))K_j^e \}
                \\
 L' &=& \mbox{} -\frac{1}{2}f^{abc} \{(\partial_j^x \delta^{bg}
                +f^{bdg}A_j^d) \delta ({\bf x} - {\bf y}) \} H_{ijl}^c
               \\
    &  &  \mbox{} -\frac{1}{4}f^{abc}F_{jk}^b \{f^{cge}(\delta_{il}B_{jk}^e
             +\delta_{jl}B_{ki}^e +\delta_{kl}B_{ij}^e)
   \delta ({\bf x} - {\bf y}) \\
            & & \mbox{} -f^{cde}
                ((\partial_i^x \delta_{jl}\delta^{dg} -\partial_j^x \delta_{il}
                \delta^{dg} +f^{dgh} \delta_{il}A_j^h +f^{dfg}\delta_{jl}
                A_i^f)\delta ({\bf x} - {\bf y}))K_k^e \\
            & & \mbox{} -f^{cde}((\partial_j^x \delta_{kl}\delta^{dg}
    -\partial_k^x
                \delta_{jl}\delta^{dg} +f^{dgh} \delta_{jl}A_k^h +f^{dfg}
                \delta_{kl}A_j^f)\delta ({\bf x} - {\bf y}))K_i^e \\
            & & \mbox{} -f^{cde}((\partial_k^x \delta_{il}\delta^{dg}
     -\partial_i^x
                \delta_{kl} \delta^{dg} +f^{dgh} \delta_{kl}A_i^h
                +f^{dfg}\delta_{il} A_k^f)\delta ({\bf x} - {\bf y}))K_j^e
                \}   \\
 M &=& f^{gac} \eta_i^c \delta ({\bf x} - {\bf y})
                \\
 M' &=& \mbox{} -f^{agc} \eta_l^c \delta ({\bf x} - {\bf y})
                \\
 N &=& \mbox{} -D_i \delta^{ag}  \delta ({\bf x} - {\bf y})
                \\
 N' &=& \mbox{} -D_l \delta^{ag}  \delta ({\bf x} - {\bf y})
                \\
 A &=& f^{gba} \Pi_{il}^b \delta ({\bf x} - {\bf y}) \\
 A' &=& \mbox{} -f^{abg} \Pi_{li}^b \delta ({\bf x} - {\bf y}) \\
 B &=& 2f^{gbc} \Pi_{li}^b (\partial_l^y \delta^{ca}
                +f^{cde}A_l^d \delta^{ea}) \delta ({\bf x} - {\bf y})
                \\
 B' &=& \mbox{} -2f^{abc} \Pi_{il}^b (\partial_i^x \delta^{cg}
                +f^{cde}A_i^d \delta^{eg}) \delta ({\bf x} - {\bf y})
                \\
 U &=& \frac{1}{4}f^{gbc} F_{mn}^b f^{cae}(F_{lm}^e
                \delta_{ni}+F_{mn}^e\delta_{li}
           +F_{nl}^e \delta_{mi}) \delta ({\bf x} - {\bf y})
                \\
 U' &=& \mbox{} -\frac{1}{4}f^{abc} F_{jk}^b f^{cge}(F_{ij}^e
                \delta_{kl}+F_{jk}^e\delta_{il}
           +F_{ki}^e \delta_{jl}) \delta ({\bf x} - {\bf y})
                \\
 V &=& (\partial_l^y \delta_{il} \delta^{ag}
               +f^{gba}A_i^b) \delta ({\bf x} - {\bf y})
                \\
 V' &=& \mbox{} -(\partial_i^x \delta_{il} \delta^{ag}
               -f^{abg}A_l^b) \delta ({\bf x} - {\bf y})
                \\
 W &=& \frac{1}{2}\delta^{ag}
                \delta ({\bf x} - {\bf y})  \\
 W' &=& \mbox{} -\frac{1}{2}\delta^{ag}
                \delta ({\bf x} - {\bf y}).
\end{eqnarray*}
After some calculation, we find the following zero mode solution
 for Eq. (\ref{m39}).
\begin{eqnarray}
\alpha_{ij} & = &(D_i \mu_j -D_j \mu_i)+2[B_{ij}, \nu]+2[F_{ij}, \xi]
            +2(D_i[\nu,K_j]- D_j[\nu,K_i]) \cr
\beta_{ij} & = & 2[\Pi_{ij} ,\nu] \cr
\gamma_i &=& 2D_i \nu  \label{f15} \\
\rho_i &=& 2[\Pi_i,\nu]  \cr
\sigma_i &=& \mu_i +2 D_i \xi \cr
\psi_i &=& 0  \cr
\phi &=& 0  
\nonumber
\end{eqnarray}
The self consistency conditions for equations
of motion, Eq. (\ref{c4}),
\[
\Omega^J \equiv (v_i^J)^T \frac{\partial V(q)}{\partial q_i} =0
\]
now vanish identically
after replacing
the above obtained zero modes:
\begin{eqnarray*}
\Omega^{(1)} &=&  \int d^3 {\bf x} \{~\alpha_{lm}^g ({\bf x}) \frac{\delta}
        {\delta B_{lm}^g ({\bf x})} +\beta_{lm}^g ({\bf x}) \frac{\delta}
      {\delta \Pi_{lm}^g ({\bf x})} +\gamma_l^g ({\bf x}) \frac{\delta}
                  {\delta A_l^g ({\bf x})}  \\
       & & \mbox{} +\rho_l^g ({\bf x}) \frac{\delta}{\delta \Pi_l^g
     ({\bf x})}
             +\sigma_l^g ({\bf x}) \frac{\delta}{\delta K_l^g ({\bf x})}
              \} \int d^3 {\bf y} V_{(1)}  \\
   & \equiv  & 0 .
\end{eqnarray*}
Thus there  are  no further constraints,
and the theory has  gauge symmetry whose symmetry transformations
are given by the above zero modes.
 Since $\gamma_i$ and $\alpha_{ij}$ in Eq. (\ref{f15}) represent
the variations of $A_i$ and $B_{ij}$ under the gauge transformation, 
respectively, we now see that both the vector and antisymmetric 
tensor fields have non-Abelian gauge symmetry
with their respective gauge parameters $\nu$ and $\mu_i$.

Now, the gauge fixing will remove the singularity completely, and
 we choose the following gauge. 
\begin{eqnarray}
\partial_i A_i =  0 , ~~~~~~~
D_i B_{ij}  =  0  \label{f16} ,
\end{eqnarray}
then the Lagrangian becomes
\begin{eqnarray}
{\cal L} &=& \frac{1}{2} \Pi_{ij}^a \dot{B}_{ij}^a + \frac{1}{2} ([A_i ,K_j]
             -[A_j ,K_i ])^a \dot{\Pi}_{ij}^a \cr
         & & \mbox{} +[A_j , \Pi_{ij}]^a \dot{K}_i^a + \frac{1}{2}
      \Pi_i^a \dot{A}_i^a
             +(D_j \Pi_{ji})^a \dot{\eta}_i^a \cr
         & & \mbox{} + (D_i \Pi_i + [B_{ij} ,\Pi_{ij}] - 2[\Pi_{ij} ,D_j K_i])^a
             \dot{\omega}^a  \label{f17} \\
         & & \mbox{} +[F_{ij} ,\Pi_{ij}]^a \dot{\theta}^a + ([\Pi_{ij} ,\Pi_j]
              -\frac{1}{4}[H_{ijk} ,F_{jk}])^a \dot{\chi}_i^a \cr
         & & \mbox{} + (D_i \eta_i^a + \frac{1}{2} \theta^a )\dot{\lambda}^a
      + (\partial_i A_i^a)\dot{\zeta}^a + (D_j B_{ij}^a)\dot{\tau}_i^a \cr
    & & \mbox{} - V_{(2)} \nonumber
\end{eqnarray}
where
\[
V_{(2)} = V_{(1)}\mid_{\{\partial_i A_i = 0 ~~,~~ D_i B_{ij} = 0 \}} .
\]
Notice that
 here we did not fix the gauge for the auxiliary vector field $K$, although
it behaves like a gauge field with the parameter $\xi$. This is because
the zero mode equation (\ref{m39}) shows that the parameters
 $\xi$ and $\mu_i$ are the
variations of $\theta$ and $\eta_i$, respectively, and
$\theta$ and $\eta_i$ are constrained by the
  reducibility condition (\ref{f8}).
Thus the gauge fixing of $B_{ij}$ does the necessary job related to 
the parameter $\xi$.
And one can check that
 the symplectic matrix obtained from the above Lagrangian is no longer
singular.

\section{Discussion and Conclusion}

In this paper, we analyze the 
symmetry of the topological mass generating action in the non-Abelian 
case with and without a vector auxiliary field.
In the Abelian case, the action which does not include
 a vector auxiliary field
develops an  effective mass for the vector gauge field 
when the
topological coupling $B \wedge F $ term is present \cite{abl}.
However, in the non-Abelian case, a straightforwardly extended action
of the Abelian type does not provide a gauge symmetry for the 
antisymmetric tensor field unless one introduces a vector auxiliary
field in a specific form.
And, if the antisymmetric tensor field does not possess a gauge symmetry,
then the physical degree of freedom of
the antisymmetric tensor field can not transmute into a component of the
vector gauge field, thus no massive vector gauge field.
Hence, it is necessary that
both the vector and antisymmetric tensor fields behave as gauge fields.
 
Recently, it was shown in Ref. \cite{hl,al} that if a vector auxiliary 
field is introduced to the action in a specific combination, then
both the vector and 
  antisymmetric tensor 
fields behave as gauge fields.

Although the action with full non-Abelian gauge symmetry was constructed 
and quantized in the BRST formalism in these works \cite{hl,al}, 
the symmetry structure related to the constraints of the theory was not
understood completely.
In this paper, we just did this remained work in the 
Faddeev-Jackiw formalism.
 
In Section II, we have shown 
that the vector field transforms as a gauge field,
but the antisymmetric tensor field does not, when there is no vector
auxiliary field:
\[ \delta A_{i} = 4 D_{i} \nu , ~~~~~ \delta B_{ij} = 4 [ B_{ij}, \nu ],
 ~~~~ ~~ {\rm etc.} \]
When we add a vector auxiliary field in Section III, 
both the vector and antisymmetric tensor fields behave
as gauge fields, that is, both transformations contain a derivative term:
\begin{eqnarray*}
\delta B_{ij} & = &(D_i \mu_j -D_j \mu_i)+2[B_{ij}, \nu]+2[F_{ij}, \xi]
            +2(D_i[\nu,K_j]- D_j[\nu,K_i]) \\
\delta A_{i} & = &  2D_i \nu \\
\delta K_{i} & = & \mu_i +2 D_i \xi \\
\vdots & & .
\end{eqnarray*}
In Ref. \cite{hl}, the transformations of fields were
given by
\begin{eqnarray}
\delta B_{\alpha \beta} & = & D_{[\alpha} \mu_{\beta ]} +
  [B_{\alpha \beta}, \nu ]
         + [ F_{\alpha \beta}, \xi ] \cr
\delta A_{\alpha} & = & D_{\alpha} \nu \cr
\delta K_{\alpha} & = & \mu_{\alpha} + D_{\alpha} \xi +
  [\nu , K_{\alpha} ]      \label{g1} \\
\vdots
\nonumber
\end{eqnarray}
where $\alpha,  \beta = 0,1,2,3$.
The two transformation laws look apparently different 
  for
the antisymmetric tensor and vector auxiliary fields.
However, 
in the action that we adopted in Section III, Eq.(\ref{f3}),
the $B$-field always appears in the combination of
$ B_{\alpha \beta} - D_{[ \alpha} K_{\beta ]} $,
and the transformation of this combined field
 is the same under both transformation rules:
\[ \delta (B_{\alpha \beta} - D_{[ \alpha} K_{\beta ]} )  =
 [ B_{\alpha \beta} - D_{[ \alpha} K_{\beta ]} , \nu] . \]
Therefore, the action has the same invariance property under
both transformation rules.
Notice that should the combined field behave as a covariant scalar,
then the auxiliary field $K$
 must behave like a gauge field.
This symmetry property was the origin of an extra scalar ghost $\kappa$
in Ref. \cite{hl}. 
In general, the antisymmetric tensor of rank two or higher
must be augmented in such a way that the augmented ones 
behave like the ordinary two 
form field strength under gauge transformation, if 
antisymmetric tensors are to behave 
as higher form gauge fields \cite{tn}.
And the above combination of the tensor field and the auxiliary vector
 field just does that work.

Finally, we turn to the issue of the reducible constraints that
 appeared in Section III when the vector auxiliary field is introduced:
Two primary constraints are related to each other by Eq. (\ref{f7}).
In order to treat these dependent constraints as independent ones, 
we introduced
another constraint expressing this fact.
Namely, we added this condition as an additional constraint, Eq.(\ref{f8}).
However, we did not use the relationship between 
the primary constraints Eq. (\ref{f7}) as a new constraint. 
Instead, we used
a relationship in which the original primary constraints were replaced by
their Lagrange multiplier fields. This is due to the fact that we
here impose the time derivative of a given constraint as a consistency
condition instead of the constraint itself.
Thus, to impose the relationship among constraints
we have to impose the constraint among their multiplier fields.
That is what we used in Eq.(\ref{f8}).
This additional condition resolves the reducibility in our case,
and we obtained the nonsingular symplectic matrix even with
a usual gauge choice in Eq.(\ref{f16}): The reducibility 
condition
also accounts for apparent lack  of gauge fixing for the 
$K$ field, since this condition also expresses that
the gauge parameter $\xi$ is related to the gauge parameter
$\mu_i$ of the field $B_{ij}$ as we explained in the
previous section.

In conclusion, introducing a vector auxiliary field 
enhanced the symmetry of the action and made both the vector
and antisymmetric tensor fields behave as gauge fields.
The reducibility of the gauge symmetry of the theory was resolved
by introducing a new constraint which properly expresses the relationship
among dependent constraints in terms of their Lagrange multiplier fields.

\acknowledgements{We would like to thank Youngjai
Kiem for helpful discussions.
This work was supported
in part by the Ministry of
Education, BSRI-96-2442, and KOSEF Grant 971-0201-007-2.
}



\end{document}